\documentclass[12pt]{article}

\usepackage{graphicx}
\usepackage{geometry}
\renewcommand{\baselinestretch}{1.3}

\usepackage{amssymb}
\usepackage{amsmath ,bm}

\newcommand{\bey}{\begin{eqnarray}}
\newcommand{\eey}{\end{eqnarray}}
\newcommand{\beq}{\begin{equation}}
\newcommand{\eeq}{\end{equation}}

\newcommand{\bs}{\boldsymbol}

\newtheorem{thm}{Theorem}[section]

\begin{document}

\vspace*{0in}

\begin{center}

{\large \bf Regression model selection via log-likelihood ratio and constrained minimum criterion}

\bigskip
Min Tsao%\footnote{CONTACT Min Tsao, mtsao@uvic.ca.}
\\{\small Department of Mathematics \& Statistics, University of Victoria, Canada}

\end{center}

\bigskip

{
\noindent {\bf Abstract:} Although the log-likelihood is widely used in model selection, the log-likelihood ratio has had few applications in this area. We develop a log-likelihood ratio based method for selecting regression models by focusing on the set of models deemed plausible by the likelihood ratio test. We show that when the sample size is large and the significance level of the test is small, there is a high probability that the smallest model in the set is the true model; thus, we select this smallest model. The significance level of the test serves as a parameter for this method. We consider three levels of this parameter in a simulation study and compare this method with the Akaike Information Criterion and Bayesian Information Criterion to demonstrate its excellent accuracy and adaptability to different sample sizes. We also apply this method to select a logistic regression model for a South African heart disease dataset.

\bigskip

}

%%%%%%%%%%%%%%%%%%%%%%%%%%%%%%%%% 1111111111111111111111111111111111111111
\section{Introduction}

Regression models are important tools for studying the relationships between response variables and predictor variables. Often, there are many predictor variables available to build a regression model but some of these variables may be inactive in the sense that they have no impact on the response. For parsimony, it is important that we identify the true model containing only and all active variables. To set up notation, consider a full regression model with $p$ predictor variables,
\beq
E(Y)=g(\mathbf{X}, \bs{\beta}),      \label{model_f}
\eeq 
where $Y$ is an $n$-vector of independent observations of the response variable, $g$ is a given function, $\mathbf{X}=[\mathbf{1}, \mathbf{x}_1,\dots,\mathbf{x}_{p}]$ is the $n\times (p+1)$ design matrix and $\boldsymbol{\beta}=(\beta_0, \beta_1,\dots,\beta_{p})^T$ is the unknown vector of regression parameters. The marginal distributions of $Y$ are assumed to be of the same type, usually in the exponential family of distributions, and known except for the values of their parameters. There may be other parameters besides $\bs{\beta}$ but they are not of interest in the context of model selection. With the above notation, a variable $\mathbf{x}_i$ is said to be active if its parameter $\beta_i\neq 0$ and inactive if $\beta_i= 0$. An important special case of (\ref{model_f}) is the generalized linear model where $E(Y)=g(\eta)$ with $\eta=\mathbf{X}\bs{\beta}$ which is the  $n$-vector of linear predictors, and $g$ is the inverse of the link function. Let ${\cal M}=\{{M}_j\}^{2^p}_{j=1}$ be the collection of $2^p$ subsets of the $p$ variables in the full model (\ref{model_f}) where each $M_j$ represents a subset. We call each $M_j$ a model as it defines a reduced model
\beq
E(Y)=g(\mathbf{X}_j, \bs{\beta}_j)   \nonumber %   \label{model_r}
\eeq
where $\mathbf{X}_j$ is the design matrix containing only variables in $M_j$ and $\bs{\beta}_j$ is the parameter vector for variables in $\mathbf{X}_j$. Throughout this paper, we adopt the classical parametric setting where ($i$) $p$ is fixed and $n>p$, and ($ii$) the full regression model is correctly specified even though its parameter values are unknown. Assumption ($ii$) implies that the true model containing only and all active variables is in ${\cal M}$.

There is a large body of literature on model selection. For a comprehensive review, see Ding, Tarokh and Yang (2018a) and Kadane and Lazar (2004).  Here, we only briefly review two commonly used model selection criteria, Akaike Information Criterion (AIC) by Akaike (1974) and Bayesian Information Criterion (BIC) by Schwarz (1978), which are part of the motivations behind the present work. The AIC approach does not assume that the underlying mechanism (true model) that generated the data is known. It selects the model from a set of candidate models that minimizes the Kullback–Leibler divergence between the fitted and the true model. Denote by $l(\hat{\bs{\beta}}_j)$ the maximum log-likelihood of model $M_j$ where  $\hat{\bm{\beta}}_j$ is the maximum likelihood estimator for $\bs{\beta}_j$, and let $d_j$ be the number of predictor variables in $M_j$. 
For selecting a model from the set ${\cal M}=\{{M}_j\}^{2^p}_{j=1}$, the AIC of a model $M_j$ is 
\beq
AIC(M_j)=-2l(\hat{\bs{\beta}}_j)+2d_j \hspace{0.2in}  \mbox{for $j=1,2,\dots,2^p$},  \label{aic}
\eeq
and the $M_j$ with the smallest AIC value is the model that minimizes the Kullback–Leibler divergence in an asymptotic sense. The AIC has an interpretation as a penalized measure of fit of $M_j$ with the fit measured by its log-likelihood and a penality term $2d_j$. For small sample situations, corrected penality terms have been proposed by several authors including Hurvich and Tsai (1989) and Broersen (2000). The BIC approach tackles the model selection problem from a Bayesian perceptive by assuming a constant prior for the models and an arbitrary prior for the parameter vector of a model. It selects the model with the largest marginal likelihood which is asymptotically equivalent to selecting the model with the minimum BIC where
\beq
BIC(M_j)=-2l(\hat{\bs{\beta}}_j)+d_j\log(n) \hspace{0.2in}  \mbox{for $j=1,2,\dots,2^p$},  \label{bic}
\eeq
which does not depend on the prior of the parameter vector. The BIC is also a penalized measure of fit. When the true model is in the set of candidate models, the BIC is consistent (Rao and Wu, 1989). Apart from AIC, BIC and their variants, there are other criteria based on penalized log-likelihood such as the Hannan and Quinn Information Criterion (Hannan and Quinn, 1979) and Bridge Criterion (Ding, Tarokh and Yang, 2018b). Indeed, the penalized log-likelihood is the most used tool to look for a model with a good balance between the fit and the size of the model. The fact that model selection methods with very different motivations have resulted in criteria with similar penalized log-likelihood forms such as (\ref{aic}) and (\ref{bic}) shows the inherent importance of the log-likelihood as a measure of fit for model selection.

The successful applications of the log-likelihood in model selection motivate us to look for a model selection strategy based on the closely related 
 log-likelihood ratio. The maximum log-likelihood ratio of a model $M_j$ is
\beq
\lambda(\hat{\bs{\beta}}_j)=-2\{ l(\hat{\bs{\beta}}_j)- l(\hat{\bs{\beta}})\}  \label{lr}
\eeq
where $\hat{\bs{\beta}}$ is the maximum likelihood estimator and $l(\hat{\bs{\beta}})$ is the maximum log-likelihood of the full model with all $p$ variables. The $\lambda(\hat{\bs{\beta}}_j)$ provides a relative measure of fit of model $M_j$ with respect to the full model. It has an important advantage over the log-likelihood $l(\hat{\bs{\beta}}_j)$ in that its value may be directly used to evaluate the plausibility of model $M_j$ because the asymptotic null distribution of  $\lambda({\bs{\beta}})$ is known to be a $\chi^2$ distribution, whereas the value of the log-likelihood of a model alone does not carry information about the plausibility of the model. To use $\lambda(\hat{\bs{\beta}}_j)$ for model selection, instead of penalizing it with a penality term proportional to the model size, we take advantage of the null distribution of $\lambda({\bs{\beta}})$ to look for a set of plausible models using the likelihood ratio test at a given significance level $\alpha$. Then, from this set of models we select the smallest model. This amounts to giving the fit of the model (as represented by the log-likelihood ratio) a higher priority over the size of the model, and minimizing the size subject to a lower bound on the fit. We refer to this approach as the constrained minimum method (CMC) for model selection. 

Tsao (2021) studied the CMC for selecting Gaussian linear models under an approximated likelihood ratio test with a significance level depending on $n$ and showed the method is consistent for selecting Gaussian linear models. The present paper uses the exact likelihood ratio test with a fixed significance level $\alpha$ and generalizes the CMC to all regression models. When $n$ is large and $\alpha$ is small, we show that there is a high probability that the smallest model in the set of plausible models is the true model. This provides an asymptotic justification for using the CMC to select regression models. In real applications, however, $n$ may not be very large and the selection of the $\alpha$ value needs to be guided by the finite sample accuracy of the resulting CMC. We will identify a default $\alpha$ value at which the CMC usually outperforms the AIC and BIC in terms of selection accuracy. Having $\alpha$ as a parameter also gives the CMC the ability to easily handle special situations such as when $n$ is small. We will discuss how to select the $\alpha$ value for such situations.

The rest of this paper is organized as follows. In Section 2, we present the CMC based on the likelihood ratio test for selecting regression models. In Section 3, we compare this method to the AIC and BIC in terms of selection accuracy in a simulation study with examples of linear, logistic and Poisson regression models. We also discuss the selection of the significance level $\alpha$. In Section 4, we apply the CMC to perform model selection for logistic regression for a South African heart disease dataset. We conclude with a few remarks in Section 5.

\section{The constrained minimum criterion}

Denote by $\bs{\beta}^t$ the true value of the regression parameter vector for the full model. Here, $\bs{\beta}^t$ is a $(p+1)$-vector and its elements corresponding to inactive variables are all zero. For simplicity, we make the following three assumptions for all regression models under consideration. The first assumption is that the maximum likelihood estimator for $\bs{\beta}^t$ based on the full model is consistent, that is, 
\beq
\hat{\bs{\beta}}\stackrel{p}{\longrightarrow}\bs{\beta}^t  \hspace{0.2in} \mbox{as $n\rightarrow \infty$}.  \label{a1}
\eeq
 The second assumption is that the null distribution of the log-likelihood ratio converges to a $\chi^2$ distribution with $p+1$ degrees of freedom, that is,
\beq
\lambda({\bs{\beta}}^t)=-2\{ l({\bs{\beta}}^t)- l(\hat{\bs{\beta}})\}  \stackrel{d}{\longrightarrow} \chi^2_{p+1}  \hspace{0.2in} \mbox{as $n\rightarrow \infty$}. \label{a2}
\eeq
By (\ref{a2}), for any fixed $\alpha \in (0,1)$, a $100(1-\alpha)\%$ asymptotic confidence region for $\bs{\beta}^t$ is
\beq
{\cal C}_\alpha=\{ \bs{\beta} \in \mathbb{R}^{p+1}: \lambda({\bs{\beta}})\leq \chi^2_{1-\alpha, p+1} \},  \label{c_region}
\eeq
where $\chi^2_{1-\alpha, p+1}$ denotes the $(1-\alpha)$th quantile of the $\chi^2_{p+1}$ distribution. The centre of this $(p+1)$-dimensional confidence region is $\hat{\bs{\beta}}$ as $\lambda(\hat{\bs{\beta}})=0$ is the smallest value of $\lambda(\bs{\beta})$. The third assumption is that the size of the confidence region ${\cal C}_\alpha$ goes to zero as $n$ goes to infinity in the sense that
\beq
\max_{\bs{\beta}\in {\cal C}_\alpha} \| \bs{\beta} -\hat{\bs{\beta}}\|_2 =o_p(1).  \label{a3}
\eeq
For commonly used regression models, regularity conditions for the asymptotic normality of the maximum likelihood estimator $\hat{\bs{\beta}}$ of the full model are available in the literature. It may be verified that assumptions  (\ref{a1}), (\ref{a2}) and  (\ref{a3}) all hold under these conditions. A stronger version of assumption (\ref{a3}), $\max_{\bs{\beta}\in {\cal C}_\alpha} \| \bs{\beta} -\hat{\bs{\beta}}\|_2 =O_p(n^{-1/2})$,
also holds under these conditions but the weaker version (\ref{a3}) is already sufficient for our subsequent discussions.
As an example of such regularity conditions, for linear regression models, a commonly used set of two such conditions are
\beq
\frac{1}{n} \sum^n_{i=1}  \mathbf{x}_{ri}\mathbf{x}_{ri}^T \rightarrow D, \nonumber %\label{cond1}
\eeq
where $\mathbf{x}_{ri}$ is the $i$th row of $\mathbf{X}$ and $D$ is a $(p+1)\times (p+1)$ positive definite matrix, and
\beq
\frac{1}{n} \max_{1\leq i \leq n} \mathbf{x}_{ri}^T\mathbf{x}_{ri} \rightarrow 0.  \nonumber % \label{cond2}
\eeq
For generalized linear models, such regularity conditions may be found in Haberman (1977), Gourieroux and Monfort (1981) and Fahrmeir and Kaufmann (1985).

The confidence region ${\cal C}_\alpha$ contains the collection of $\bs{\beta} \in \mathbb{R}^{p+1}$ not rejected by the likelihood ratio test for $H_0: \bs{\beta}^t=\bs{\beta}$ at the given $\alpha$ level. As such, it represents the set of plausible $(p+1)$-vectors at the $\alpha$ level. To extend the notion of plausibility from a $(p+1)$-vector to a model $M_j$, we first find a $(p+1)$-vector to represent model $M_j$. The maximum likelihood estimator $\hat{\bs{\beta}}_j$ for $M_j$ is a vector of dimension $(d_j+1)$, which is less than $(p+1)$ when $M_j$ is not the full model. It is usually a continuous random vector, so with probability one none of its elements is zero. We augment the dimension of $\hat{\bs{\beta}}_j$ by adding $(p-d_j)$ zeros to its elements to represent the $(p-d_j)$ variables not in $M_j$. For example, if $\mathbf{x}_1$ is not in $M_j$, then the second element of the augmented $\hat{\bs{\beta}}_j$ (which corresponds to $\mathbf{x}_1$) is a zero. For simplicity, we still use the same notation $\hat{\bs{\beta}}_j$ but it is now a $(p+1)$-vector representing $M_j$ and its non-zero elements correspond to the intercept and variables in $M_j$. We say that model $M_j$ is plausible at the $\alpha$ level if $\hat{\bs{\beta}}_j$ is in the confidence region ${\cal C}_\alpha$. Alternatively, we may also say that $M_j$ is plausible if $\lambda(\hat{\bs{\beta}}_j)$ is less than $\chi^2_{1-\alpha, p+1}$. Note that although we need the augmented  $(p+1)$-dimensional version of $\hat{\bs{\beta}}_j$ to define the plausibility of its corresponding model $M_j$, when computing the maximum log-likelihood ratio of this model $\lambda(\hat{\bs{\beta}}_j)$, $\hat{\bs{\beta}}_j$ may be either the augmented  $(p+1)$-dimensional version or the original $(d_j+1)$-dimensional version as they both give the same value of $\lambda(\hat{\bs{\beta}}_j)$. In numerical computations of $\lambda(\hat{\bs{\beta}}_j)$, we use the $(d_j+1)$-dimensional version as it appears in (\ref{lr}) which is more convenient. Using the $L_0$ norm which counts the number of non-zero elements in a vector, we define the constrained minimum criterion based on the likelihood ratio test as the criterion that chooses the model represented by the solution of the following constrained minimization problem
\beq
\underset{\bm{{\cal M}}}{\text{minimize}}       \|\hat{\bm{\beta}}_j\|_0
\mbox{\hspace{0.1in} subject to \hspace{0.01in} } \hat{\bm{\beta}}_j \in {\cal C}_\alpha.  \label{form1}
\eeq
We call the solution vector to this minimization problem the CMC solution and its corresponding model the CMC selection. When there are multiple solution vectors, we choose the one with the highest likelihood as the CMC solution.

Denote by $\hat{\bm{\beta}}_{j}^t$ the maximum likelihood estimator for the unknown true model $M^t_j$. Non-zero elements of this (augmented) $\hat{\bm{\beta}}_{j}^t$ are those corresponding to active variables and zero elements are those corresponding to inactive variables. The following theorem gives the asymptotic properties of the CMC solution and selection.

% Theorem Theorem Theorem ---------------------------------------------------------------------------------------------------------------
\begin{thm} \label{thm1}

Suppose assumptions (\ref{a1}),  (\ref{a2}) and  (\ref{a3}) hold. For a given $\alpha \in (0,1)$, 
let $\hat{\bm{\beta}}_{\alpha}$ be the CMC solution of (\ref{form1}) and $\hat{M}_\alpha$ be the corresponding CMC selection. Then, (i) the CMC solution is consistent in that
\beq \hat{\bm{\beta}}_{\alpha} \stackrel{p}{\longrightarrow} \bm{\beta}^t \hspace{0.2in} \mbox{as $n\rightarrow \infty$}, \label{convg1} \eeq
and (ii) the probability that $\hat{M}_\alpha$ is the true model has an asymptotic lower bound
\beq \lim_{n\rightarrow +\infty}P(\hat{M}_{\alpha} = M_j^t) \geq 1-\alpha. \label{convg2} \eeq

\end{thm}

\vspace{0.1in}

The asymptotic lower bound (\ref{convg2}) shows that when the sample size $n$ is large, we may choose a small $\alpha$ so that there is a high probability that the CMC selection is the true model. Numerical results show that the lower bound $(1-\alpha)$ is rather loose for many $\alpha$ values when $n$ is large in that the observed probability of the event $\{\hat{M}_{\alpha} = M_j^t\}$ is usually much larger than $(1-\alpha)$. Also, when $n$ is not large, small $\alpha$ levels are not appropriate. We will discuss the selection of the $\alpha$ level with numerical examples in the next section. We now prove the theorem.

\vspace{0.1in}

\noindent {\bf Proof of Theorem \ref{thm1}.} 
By (\ref{a1}), we have  $\|\hat{\bm{\beta}}-\bm{\beta}^t\|_2 =o_p(1)$. Since $\hat{\bm{\beta}}_{\alpha} \in {\cal C_\alpha}$, by (\ref{a3}) we also have  $\|\hat{\bm{\beta}}_{\alpha}-\hat{\bm{\beta}}\|_2=o_p(1)$. It follows from these and the triangle inequality that
\beq
\|\hat{\bm{\beta}}_{\alpha}-\bm{\beta}^t\|_2\leq 
\|\hat{\bm{\beta}}_{\alpha}-\hat{\bm{\beta}}\|_2 + \|\hat{\bm{\beta}}-\bm{\beta}^t\|_2 =o_p(1),
\label{op1}
\eeq
which implies the consistency of the CMC solution (\ref{convg1}).

To prove the asymptotic lower bound in (\ref{convg2}), note that
\beq 
P(\hat{M}_{\alpha} = M_j^t)=P(\hat{\bm{\beta}}_{\alpha} = \hat{\bm{\beta}}_{j}^t)  \label{equal}
\eeq
as events $\{\hat{M}_{\alpha} = M_j^t\} \Longleftrightarrow \{\hat{\bm{\beta}}_{\alpha} = \hat{\bm{\beta}}_{j}^t\}$, 
so it suffices to show that $P(\hat{\bm{\beta}}_{\alpha} = \hat{\bm{\beta}}_{j}^t)$ has the asymptotic lower bound in  (\ref{convg2}). To this end, we
first identify the elements of vectors $\bm{\beta}$ in ${\cal C}_{\alpha}$ that may not be zero when $n$ is large. Define an event
\[ \mbox{
$A=$ \{Elements of $\bm{\beta}$ in ${\cal C}_{\alpha}$ corresponding to non-zero elements of $\bm{\beta}^t$ are also non-zero\}. }
\]
Similar to (\ref{op1}), by the triangle inequality and (\ref{a3}),  we have $\|{\bm{\beta}}-\bm{\beta}^t\|_2 =o_p(1)$ uniformly for all  $\bm{\beta} \in {\cal C}_{\alpha}$. It follows that individual elements of $\bm{\beta}$ converge in probability to corresponding elements of $\bm{\beta}^t$ uniformly, so $P(A) \rightarrow 1$  as the sample size $n$ goes to infinity. When event $A$ occurs, among the set of vectors $\{{\hat{\bm{\beta}}_j} \}_{j=1}^{2^p}$ only those for models containing all active variables can be in ${\cal C}_{\alpha}$, so $\hat{\bm{\beta}}_{j}^t$ of the true model $M^t_j$ is the smallest (in $L_0$ norm) member of $\{{\hat{\bm{\beta}}_j} \}_{j=1}^{2^p}$ that may possibly be in ${\cal C}_{\alpha}$. It follows that $\{\hat{\bm{\beta}}_{j}^t \in {\cal C}_{\alpha}\}\cap {A}$ implies $\{\hat{\bm{\beta}}_{\alpha} = \hat{\bm{\beta}}_{j}^t\}$, so
\beq
P(\hat{\bm{\beta}}_{\alpha} = \hat{\bm{\beta}}_{j}^t) \geq
P( \{\hat{\bm{\beta}}_{j}^t \in {\cal C}_{\alpha}\} \cap {A}) \rightarrow P( \hat{\bm{\beta}}_{j}^t \in {\cal C}_{\alpha})  \label{temp1}
\eeq
as $n$ goes to infinity. Also, event $\{{\bm{\beta}^t}\in {\cal C}_{\alpha}\}$ implies $\{\hat{\bm{\beta}}_{j}^t \in {\cal C}_{\alpha}\}$ because $\hat{\bm{\beta}}_{j}^t$ is the maximum likelihood estimator for $M_j^t$ which has a higher likelihood and thus a smaller log-likelihood ratio than ${\bm{\beta}^t}$; that is, $\lambda(\hat{\bm{\beta}}_{j}^t)<\lambda({\bm{\beta}}^t)$ and thus $\{{\bm{\beta}^t}\in {\cal C}_{\alpha}\}$ implies $\{\hat{\bm{\beta}}_{j}^t \in {\cal C}_{\alpha}\}$. This and (\ref{a2}) imply that
\beq
P( \hat{\bm{\beta}}_{j}^t \in {\cal C}_{\alpha})\geq P(\bm{\beta}^t \in {\cal C}_{\alpha}) \rightarrow 1-\alpha  \label{temp2}
\eeq
as $n$ goes to infinity. Equations (\ref{equal}),  (\ref{temp1}) and (\ref{temp2}) then imply (\ref{convg2}).  \hfill $\Box$

\vspace{0.1in}

The above proof follows similar steps as the proof of the consistency of the CMC for Gaussian linear models in Tsao (2021). However, that CMC for Gaussian linear models is based on an approximated likelihood ratio statistic whose finite sample distribution is known. Its $\alpha$ level is not fixed and goes to zero as $n$ goes to infinity. The proof of its consistency depends on the finite sample distribution and a decreasing $\alpha$. In the present paper, the $\alpha$ is fixed and the finite sample distribution of the likelihood ratio statistic is unavailable. We have only the asymptotic distribution (\ref{a2}) which leads to a weaker result (\ref{convg2}) instead of consistency. 
Nevertheless, this does not seem to affect the accuracy of the present version of CMC based on the exact likelihood ratio test as numerical results show that it is equally accurate as the consistent version for Gaussian linear models (see numerical examples in the next section). Further, the present version can be applied to all types of regression models satisfying the three assumptions, not just the Gaussian linear models.

\section{Simulation study}

We now compare the CMC based on the likelihood ratio test (\ref{form1}) with the AIC and BIC in terms of false active rate (FAR) and false inactive rate (FIR) through numerical examples. We also discuss the selection of the $\alpha$ level for the CMC. Here, FAR is the number of inactive variables appearing in the selected model divided by the total number of inactive variables in the full model, and FIR is the number of active variables not in the selected model divided by the total number of active variables in the full model. A model selection criterion is accurate when FIR and FAR of its selected model are both low.
%It may be verified that the three assumptions of the last section hold for all examples in this section, so Theorem \ref{thm1} applies. 
To compute the examples, we use R package `bestglm' by McLeod, Xu and Lai (2020) which performs the best subset selection for generalized linear models. For the best subset selection of Gaussian linear models, `bestglm' uses the `leaps and bounds' algorithm by Furnival and Wilson (1974) which can handle situations with 40 or fewer predictor variables. For the best subset selection of logistic regression models and Poisson regression models, it uses a complete enumeration method  by Morgan and Tatar (1972) and has a limit of 15 on the number of predictor variables allowed in the full model. In our simulation examples, we set the number of predictor variables below the limit to a maximum of 30 for linear models and 10 for the two generalized linear models to avoid long simulation time.

%The performance of the model selection criteria depends on the parameter values of the underlying true model and the sample size. For all examples, the parameter values and minimum sample size values are chosen so that model selection is a meaningful exercise in the sense that the maximum of FIR and FAR is below 50\% for all methods. 

\subsection{Linear model examples}

The linear model used for comparison is
\begin{equation}
\mathbf{y}=\mathbf{X}\bm{\beta} +\bm{\varepsilon}, \label{m01}
\end{equation}
where $\boldsymbol{\varepsilon} \sim N(\mathbf{0}, \sigma^2\mathbf{I})$ with $\sigma^2=1$, $\mathbf{X}=[\mathbf{1}, \mathbf{x}_1,\dots,\mathbf{x}_{p}]$, and $\boldsymbol{\beta}=(1, \beta_1,\dots,\beta_{p^*}, 0,\dots, 0)^T$ with $\beta_1=\dots=\beta_{p^*}=1$, so only the first $p^*$ variables are active. Elements of all $\mathbf{x}_i$ are independent random numbers generated from the standard normal distribution. 
The performance of the CMC depends on the $\alpha$ level. To find the appropriate levels for different sample sizes, we consider three levels, $\alpha=0.1, 0.5$ and 0.9. 

Table 1 contains simulated values of the (FIR, FAR) pairs for five model selection criteria, AIC, BIC, CMC$_{0.9}$, CMC$_{0.5}$ and CMC$_{0.1}$, at 12 different combinations of $n$, $p$ and $p^*$.  The subscript $\alpha$ in CMC$_{\alpha}$ indicates the $\alpha$ level used. Each  (FIR, FAR) pair in the table is based on 1000 simulation runs. For each run, we first generate an ($\mathbf{X}, \mathbf{y}$) pair, and then perform the best subset selection using  ($\mathbf{X}, \mathbf{y}$) with the five criteria to find their chosen models and compute their (FIR, FAR) values based on their chosen models. After 1000 runs, we obtain 1000 (FIR, FAR) values for each criteria, and Table 1 contains the average of these 1000 values. We make the following comments based on results in Table 1.

\begin{table}
\caption{\label{tb-1} Model selection accuracy comparison for Gaussian linear models: the entries are simulated (FIR, FAR) of the AIC, BIC and three CMC criteria for 12 scenarios. The bold  CMC results are those at the recommended  $\alpha$ level. Results in the table are the average of the two rates for 1000 simulation runs rounded to the second digit after the decimal point.}
\centering
{\small
\begin{tabular}{lccccc} \\
$(n, \hspace{0.03in} p, \hspace{0.05in} p^*)$ & \textsc{aic} & \textsc{bic} &\textsc{cmc}$_{0.9}$ & \textsc{cmc}$_{0.5}$ & \textsc{cmc}$_{0.1}$ \\ \hline %\\[0.01pt]
(20, 10, 5)   & (0.04, 0.34) & (0.05, 0.24) & (0.05, 0.25)  & {\bf (0.09, 0.13)} & (0.21, 0.06) \\
(30, 10, 5)   & (0.00, 0.25) & (0.01, 0.12) & (0.01, 0.16) & {\bf (0.02, 0.05)} & (0.09, 0.01)\\
(40, 10, 5)   & (0.00, 0.24) & (0.00, 0.09) & (0.00, 0.13) & {\bf (0.00, 0.04)} & {(0.03, 0.01)}\\
(50, 10, 5)   & (0.00, 0.22) & (0.00, 0.08) & (0.00, 0.12) & {\bf (0.00, 0.03)} & {(0.01, 0.00)}\\  \hline %\\[0.01pt]

(40, 20, 10)   & (0.00, 0.32) & (0.00, 0.15) & (0.01, 0.12) & {\bf (0.02, 0.05)} & (0.06, 0.02)\\
(60, 20, 10)   & (0.00, 0.25) & (0.00, 0.09) & (0.00, 0.08) & {\bf (0.00, 0.02)} & (0.01, 0.00)\\
(80, 20, 10)   & (0.00, 0.21) & (0.00, 0.06) & (0.00, 0.05) & {\bf (0.00, 0.01)} & {(0.00, 0.00)}\\
(100, 20, 10)   & (0.00, 0.20) & (0.00, 0.05) & (0.00, 0.05) & {\bf (0.00, 0.01)} & {(0.00, 0.00)} \\  \hline %\\[0.01pt]

(60, 30, 15)   & (0.00, 0.31) & (0.00, 0.12) & (0.00, 0.08) & {\bf (0.00, 0.03)} & (0.02, 0.01)\\
(90, 30, 15)   & (0.00, 0.23) & (0.00, 0.07) & (0.00, 0.04) & {\bf (0.01, 0.01)} & (0.01, 0.00)\\
(120, 30, 15)   & (0.00, 0.21) & (0.00, 0.05) & (0.00, 0.03) & {\bf (0.00, 0.00)} & {(0.00, 0.00)}\\
(150, 30, 15)   & (0.00, 0.20) & (0.00, 0.04) & (0.00, 0.02) & {\bf (0.00, 0.00)} & {(0.00, 0.00)}

\end{tabular}
}
\end{table}

\begin{itemize}

\item[1.] The AIC and BIC have low FIR, but the AIC has a high FAR of more than 20\% even when the sample size $n$ is five times as large as the dimension $p$. If we treat false active and false inactive as equally serious errors and rank the five criteria by the overall error rate defined as the sum of FIR and FAR, then the AIC has the highest overall error rate regardless the dimension and sample size. The BIC is consistent, and we see that its overall error rate is going down towards zero as the sample size increases.
\item[2.] The performance of the CMC$_{0.9}$ is similar to that of the BIC with comparable FIR and FAR. For small and moderate sample sizes of $n\leq 3p$, the CMC$_{0.5}$ has in general the smallest overall error rate among the five criteria. For large sample sizes of $n>3p$, the CMC$_{0.1}$ has the smallest overall error rate but CMC$_{0.5}$ is a close second. Because of these, we recommend the 0.5 level as the default $\alpha$ level for the CMC. CMC results at this recommanded default level are highlighted in bold fonts in Table 1, and they are substantially more accurate than that of the AIC and BIC. When the sample size $n$ is very large relative to the dimension $p$, we may use the 0.1 level.
\item[3.] Although the three assumptions in the previous section were insufficient for proving the consistency of the present version of the CMC based on the likelihood ratio test, Table 1 shows when $n$ is large and $\alpha$ is small, the CMC overall error rates are zero or very close to zero. This suggests that for Gaussian linear models, the present version of the CMC is also consistent when we let the $\alpha$ level go to zero at a certain speed as the sample size increases. Further, comparing the CMC$_{0.1}$ results with the BIC results, we see that the CMC selection appears to converge to the true model faster than the BIC selection as the BIC error rates never reached zero even when $n=5p$.

\end{itemize}

Model (\ref{m01}) was also used to evaluate the consistent CMC for Gaussian linear models in Table 1 of Tsao (2021).
The CMC results in Table 1 of that paper differ from the CMC results in Table 1 here, especially for the small sample cases of $n=2p$. These differences are due to the fact that two different tests were used in the formulation of the CMC. The tests are asymptotically equivalent, so for large sample sizes ($n>3p$) the CMC results in both tables are very similar. In the examples reported here, we had set $p^*=p/2$ so that there is an equal number of active and inactive variables which makes the use of FIR+FAR as a measure of the overall error the most meaningful. For simplicity, we also set parameters of all active variables to 1. We have tried other $p^*$ values and parameter values, and obtained similar observations concerning the relative performance of the five criteria.

\subsection{Logistic regression examples}

Let $Y_1, Y_2, \dots, Y_n$ be $n$ independent observations of the response variable where $Y_i \sim Binomial(m, \pi_i)$ and let $\mathbf{X}=[\mathbf{1}, \mathbf{x}_1,\dots,\mathbf{x}_{p}]$ be the corresponding $n\times (p+1)$ matrix of predictor variables. The logistic regression model is given by
\beq
	logit(\pi_i)=\mathbf{x}_{ri}\bs{\beta},  \label{logit}
\eeq
or alternatively,
\beq
	\pi_i=\frac{\exp(\mathbf{x}_{ri}\bs{\beta})}{1+ \exp(\mathbf{x}_{ri}\bs{\beta})} , \nonumber  %\label{logistic}
\eeq
where $\mathbf{x}_{ri}$ is the $i$th row of $\mathbf{X}$ and $\boldsymbol{\beta}=(1, \beta_1,\dots,\beta_{p^*}, 0,\dots, 0)^T$. As in the linear model examples, we set $\beta_1=\dots=\beta_{p^*}=1$ and $p^*=p/2$, so that only the first half of the variables are active, and elements of all predictor variables $\mathbf{x}_i$ are independent random numbers generated from the standard normal distribution. The sample size here depends on both $n$ and $m$, so we used different combinations of $n$ and $m$ in the simulation. Table 2 contains the (FIR, FAR) values of the AIC, BIC, CMC$_{0.9}$, CMC$_{0.5}$ and CMC$_{0.1}$ for 16 combinations of $(n, m, p)$ where each (FIR, FAR) is the average of 1000 simulated pairs. We make the following observations based on Table 2:

\begin{table}
\caption{\label{tb-2} Model selection accuracy comparison for logistic regression models: the entries are simulated (FIR, FAR) of the AIC, BIC and three CMC criteria for 16 scenarios. The bold CMC results are those at the  recommended  $\alpha$ level. Results in the table are the average of the two rates for 1000 simulation runs rounded to the second digit after the decimal point.}
\centering
{\small
\begin{tabular}{lccccc} \\
$(n, \hspace{0.03in} m, \hspace{0.03in} p, \hspace{0.03in} p^*)$ & \textsc{aic} & \textsc{bic} &\textsc{cmc}$_{0.9}$ & \textsc{cmc}$_{0.5}$ & \textsc{cmc}$_{0.1}$ \\ \hline %\\[0.01pt]
(20, 5, 6, 3)   & (0.06, 0.20) & (0.10, 0.11) & (0.06, 0.20)  & {\bf (0.14, 0.07)} & (0.30, 0.03) \\
(30, 5, 6, 3)   & (0.01, 0.17) & (0.02, 0.08) & (0.01, 0.17) & {\bf (0.03, 0.05)} & (0.12, 0.01)\\
(40, 5, 6, 3)   & (0.00, 0.17) & (0.00, 0.07) & (0.00, 0.16) &  {\bf (0.01, 0.04)} & { (0.03, 0.00)}\\
(50, 5, 6, 3)   & (0.00, 0.16) & (0.00, 0.06) & (0.00, 0.16) &  {\bf (0.00, 0.04)} & { (0.01, 0.00)}\\  \hline %\\[0.01pt]

(20, 10, 6, 3)   & (0.01, 0.17) & (0.01, 0.10) & (0.01, 0.17)  & {\bf (0.02, 0.05)} & (0.08, 0.01) \\
(30, 10, 6, 3)   & (0.00, 0.16) & (0.00, 0.07) & (0.00, 0.16) & {\bf (0.00, 0.04)} & (0.01, 0.00)\\
(40, 10, 6, 3)   & (0.00, 0.16) & (0.00, 0.06) & (0.00, 0.16) & {\bf (0.00, 0.04)} & { (0.00, 0.00)}\\
(50, 10, 6, 3)   & (0.00, 0.16) & (0.00, 0.05) & (0.00, 0.16) & {\bf (0.00, 0.03)} & {(0.00, 0.00)}\\  \hline %\\[0.01pt]

(20, 5, 10, 5)   & (0.17, 0.26) & (0.22, 0.17) & (0.19, 0.20)  & {\bf (0.30, 0.11)} & (0.42, 0.06) \\
(30, 5, 10, 5)   & (0.04, 0.19) & (0.07, 0.10) & (0.06, 0.13) & {\bf (0.12, 0.05)} & (0.24, 0.02)\\
(40, 5, 10, 5)   & (0.00, 0.18) & (0.02, 0.07) & (0.01, 0.10) & {\bf (0.04, 0.03)} & {(0.13, 0.01)}\\
(50, 5, 10, 5)   & (0.00, 0.16) & (0.00, 0.05) & (0.00, 0.08) & {\bf (0.01, 0.02)} & {(0.07, 0.00)}\\  \hline %\\[0.01pt]

(20, 10, 10, 5)   & (0.06, 0.19) & (0.08, 0.11) & (0.08, 0.13)  & {\bf (0.15, 0.06)} & (0.26, 0.04) \\
(30, 10, 10, 5)   & (0.00, 0.17) & (0.01, 0.07) & (0.00, 0.10) & {\bf (0.01, 0.02)} & (0.06, 0.01)\\
(40, 10, 10, 5)   & (0.00, 0.15) & (0.00, 0.06) & (0.00, 0.07) & {\bf (0.00, 0.02)} & {(0.02, 0.00)}\\
(50, 10, 10, 5)   & (0.00, 0.14) & (0.00, 0.05) & (0.00, 0.07) & {\bf (0.00, 0.02)} & {(0.00, 0.00)}\\  \hline %\\[0.01pt]

\end{tabular}
}
\end{table}

\begin{itemize}

\item[1.] The AIC has the lowest FIR but the highest FAR  for all combinations of $(n,m,p,p^*)$. Due to its high FAR, its overall error rate is in general the highest among the five criteria. The BIC has much lower FAR than the AIC. It is consistent but its FAR converges to zero slowly as $n$ and $m$ increase, and it is still about 5\% even when $n$ and $m$ are at their highest values of 50 and 10, respectively.
\item[2.] We had noted that for Gaussian linear models, the performance of the CMC$_{0.9}$ is similar to that of the BIC with low FIR. However, for logistic regression models, CMC$_{0.9}$ behaves more like the AIC with similar FIR and FAR, especially for cases where $(p,p^*)=(6,3)$. When $(p,p^*)=(10,5)$, it has smaller FAR than the AIC. 
\item[3.] For Gaussian linear models, we have recommended the 0.5 level as the default $\alpha$ level for the CMC. For logistic regression model selection, both $n$ and $m$ affect the accuracy of the CMC. Interestingly, however, through exploring a wide range of $n$ and $m$ combinations we found that CMC$_{0.5}$ again has a stable performance and is usually the most or the second most accurate criterion among the five criteria. We thus also recommend the $0.5$ level as the default level for logistic regression model selection. When $n\times m$ is much larger than $p$, CMC$_{0.1}$ may be used instead. Table 2 has such cases where $n\times m$ is 50 times as large as $p$, and for these cases CMC$_{0.1}$ reached zero error rates, suggesting that the CMC is also consistent for selecting logistic regression models.

\end{itemize}

\subsection{Poisson regression examples}

Let $Y_1, Y_2, \dots, Y_n$ be $n$ independent observations of the response variable where $Y_i \sim Poisson(\mu_i)$. The Poisson regression model with log link is
\beq
	\ln(\mu_i)=\mathbf{x}_{ri}\bs{\beta} \label{Poisson}
\eeq
where $\mathbf{x}_{ri}$ is the $i$th row of the $n\times (p+1)$ matrix $\mathbf{X}=[\mathbf{1}, \mathbf{x}_1,\dots,\mathbf{x}_{p}]$ of predictor variables and $\bs{\beta}$ is the vector of regression parameters. We set $\boldsymbol{\beta}=(1, \beta_1,\dots,\beta_{p^*}, 0,\dots, 0)^T$ with $p^*=p/2$ and $\beta_1=\dots=\beta_{p^*}=0.5$. Table 3 contains the simulated (FIR, FAR) results for the 5 criteria. Like the case of logistic regression models in Table 2, the performance of the CMC$_{0.9}$ is similar to that of the AIC which has the lowest FIR but the highest FAR.  The BIC has slightly higher FIR than the AIC and CMC$_{0.9}$ but lower FAR. On the relative performance of the three CMC criteria, CMC$_{0.9}$ has lower overall error when the sample size $n$ is small ($n\leq 2p$). When $2p<n\leq 4p$, CMC$_{0.5}$ is usually the most accurate. When $n>4p$, CMC$_{0.1}$ is usually the most accurate  but CMC$_{0.5}$ is a close second. Based on these findings and for simplicity, we again recommend the 0.5 level as the default level. For small sample sizes, the 0.9 level may be used. For very large sample sizes, the 0.1 level may be used.

\begin{table}
\caption{\label{tb-3} Model selection accuracy comparison for Poisson regression models: the entries are simulated (FIR, FAR) of the AIC, BIC and three CMC criteria for 10 scenarios. The bold  CMC results are those at the  recommended  $\alpha$ level. Results in the table are the average of the two rates for 1000 simulation runs rounded to the second digit after the decimal point.}
\centering
{\small
\begin{tabular}{lccccc} \\
$(n, \hspace{0.03in} p, \hspace{0.03in} p^*)$ & \textsc{aic} & \textsc{bic} &\textsc{cmc}$_{0.9}$ & \textsc{cmc}$_{0.5}$ & \textsc{cmc}$_{0.1}$ \\ \hline %\\[0.01pt]
(20, 6, 3)   & (0.06, 0.19) & (0.09, 0.10) & (0.05, 0.20)  & {\bf (0.12, 0.06)} & (0.28, 0.03) \\
(30, 6, 3)   & (0.01, 0.16) & (0.02, 0.07) & (0.01, 0.16) & {\bf (0.02, 0.03)} & (0.08, 0.01)\\
(40, 6, 3)   & (0.00, 0.16) & (0.01, 0.06) & (0.00, 0.17)  & {\bf (0.01, 0.03)} & {(0.03, 0.00)}\\
(50, 6, 3)   & (0.00, 0.15) & (0.00, 0.05) & (0.00, 0.16) &  {\bf (0.00, 0.03)} & {(0.01, 0.00)}\\ 
(100, 6, 3)   & (0.00, 0.15) & (0.00, 0.03) & (0.00, 0.15) & {\bf (0.00, 0.02)} & {(0.00, 0.00)}\\  \hline %\\[0.01pt]

(20, 10, 5)   & (0.13, 0.20) & (0.17, 0.14) & {(0.16, 0.15)}  & {\bf (0.26, 0.08)} & (0.39, 0.06) \\
(30, 10, 5)   & (0.01, 0.16) & (0.03, 0.07) & (0.02, 0.09) & {\bf (0.06, 0.03)} & (0.15, 0.02)\\
(40, 10, 5)   & (0.00, 0.16) & (0.00, 0.06) & (0.00, 0.08) & {\bf (0.01, 0.02)} & {(0.05, 0.00)}\\
(50, 10, 5)   & (0.00, 0.16) & (0.00, 0.05) & (0.00, 0.08) & {\bf (0.00, 0.01)} & {(0.01, 0.00)}\\
(100, 10, 5)   & (0.00, 0.16) & (0.00, 0.03) & (0.00, 0.07) & {\bf (0.00, 0.01)} & {(0.00, 0.00)}\\  \hline %\\[0.01pt]

\end{tabular}
}
\end{table}

To summarize the simulation study,
the recommendations on the $\alpha$ level that we have made in this section are based on the objective of minimizing the overall error. For fixed $n$ and $p$, the FIR of the CMC decreases and the FAR increases when $\alpha$ increases. This gives users of the CMC control on the balance between these two rates through the choice of the $\alpha$ level. If a low FAR is the priority instead of a lower overall error, one can set $\alpha$ to 0.1 regardless the sample size $n$ and dimension $p$. If a low FIR is the priority, one can set it to 0.9. We have only considered three $\alpha$ levels here. Other levels may also be used. For example, in Table 1 for the linear model (\ref{m01}), the lowest $\alpha$ level is $0.1$. When $n\gg p$ such as $(n, p)=(200, 20)$, even smaller $\alpha$ levels such as 0.05 may be used (we tried CMC$_{0.05}$ for this case and obtained zero error rates). Finally, we note that predictor variables in the above examples have low correlations as they are independently generated. When there are strongly correlated predictor variables, simulation results (not included here) show that CMC$_{0.9}$ may be more accurate than CMC$_{0.5}$ and CMC$_{0.1}$ for small and moderate sample sizes. Nevertheless, CMC$_{0.5}$ is still often the most or second most accurate, and is often substantially more accurate than the AIC and BIC. Because of these, we recommend the 0.5 level as the default regardless the type of regression model, the sample size and the correlation situation of the predictor variables. This makes the application of the CMC straightforward as a user does not have to spend time deciding on which $\alpha$ level to use. However, to optimize the CMC, one may consider a different level depending on the sample size and correlation situation.

\begin{table}
\caption{\label{tb-4} Estimated full logistic regression model for the South Africa heart disease data}
\centering
\begin{tabular}{ccccc} \\

Variable  & Estimate    & Std. Error & $z$ value  & $p$-value  \\ \hline
(Intercept) &   -6.1507208650& 1.308260018& -4.70145138 &2.583188e-06 \\
sbp            & 0.0065040171 &0.005730398 & 1.13500273 &2.563742e-01\\
tob        & 0.0793764457 &0.026602843&  2.98375801 &2.847319e-03\\
ldl            & 0.1739238981& 0.059661738 & 2.91516648 &3.554989e-03\\
adi    &   0.0185865682 &0.029289409 & 0.63458325 &5.257003e-01\\
fhd  &0.9253704194 &0.227894010 & 4.06052980& 4.896149e-05\\
typ         &0.0395950250 &0.012320227&  3.21382267 &1.309805e-03\\
obe        &-0.0629098693 &0.044247743 &-1.42176449 &1.550946e-01\\
alc        &0.0001216624 &0.004483218 & 0.02713729 &9.783502e-01\\
age             &0.0452253496 &0.012129752 & 3.72846442 & 1.926501e-04\\

\end{tabular}
\end{table}

\section{South African heart disease data analysis}

We now apply the CMC to perform model selection for logistic regression for a dataset from a heart disease study conducted by Rousseauw et al. (1983). The dataset can be found in various publicly available sources such as the R package `bestglm' by McLeod, Xu and Lai (2020) and the online resource for the book {\em Elements of Statistical Learning} by Hastie, Tibshirani and Friedman (2009). The response variable in the dataset is the coronary heart disease status (chd), a binary variable recording the presence (chd=1) or absence (chd=0) of coronary heart disease for a sample of 462 males from a heart disease high risk region of the Western Cape, South Africa. There are 9 predictor variables: systolic blood pressure (sbp), tobacco use (tob), low density lipoprotein cholesterol (ldl), adiposity (adi), family history of heart disease (fhd), type-A behavior (typ), obesity (obe), alcohol consumption (alc), age at onset (age). Fitting the full logistic regression model to chd using all 9 predictor variables yields the output in Table \ref{tb-4}. Five variables have small $p$-values, and in ascending order of their $p$-values these 5 variables are fhd, age, typ,  tob, and ldl.

\begin{table}
\caption{\label{tb-5} Models with the highest likelihood. Each row represents the model with the highest likelihood among models with the same number of variables. ``1" indicates the variable in the column heading is in the model, and ``0" means it is not in the model. The ``LogLR" column gives the maximum log-likelihood ratios of the models. Symbol $^\ddag$ indicates the model chosen by AIC, BIC, CMC$_{0.9}$ and CMC$_{0.5}$, and $^\dag$ indicates the model chosen by  CMC$_{0.1}$. }
\centering
\begin{tabular}{cccccccccccc} \\

sdp & tob & ldl & adi & fhd & typ & obe & alc & age & AIC & BIC & LogLR\\ \hline
0    &  0   &  0 &  0   & 0    & 0   &  0    &  0  &  0 &  596.1084 & 596.1084 & 123.96\\
0    &  0   &  0 &  0   & 0    & 0   &  0    &  0  &  1 &  527.5623 & 531.6979 & 53.422\\
0    &  0   &  0 &  0   & 1    & 0   &  0    &  0  &  1 &  510.6582 & 518.9293 & 34.518\\
0    &  1   &  0 &  0   & 1    & 0   &  0    &  0  &  1  & 501.3854 & 513.7921 & 23.245\\
0    &  1   &  0 &  0   & 1    & 1   &  0    &  0  &  1  & 492.7143 & 509.2566 & 12.574$^\dag$\\
0    &  1   &  1 &  0   & 1    & 1   &  0    &  0  &  1 & 485.6856$^\ddag$  & 506.3634$^\ddag$ & 3.5455$^\ddag$\\
0    &  1   &  1 &  0   & 1    & 1   &  1    &  0  &  1 & 485.9799  & 510.7933 & 1.8398\\
1    &  1   &  1 &  0   & 1    & 1   &  1    &  0  &  1  & 486.5490 & 515.4979 & 0.4089\\
1    &  1   &  1 &  1   & 1    & 1   &  1    &  0  &  1 & 488.1408 & 521.2253 & 0.0001\\
1    &  1   &  1 &  1   & 1    & 1   &  1    &  1  &  1 & 490.1400 & 527.3601 & 0.0000\\

\end{tabular}
\end{table}

Using `bestglm', we obtain the 10 models with the highest likelihood among models with the same number of predictor variables. These 10 models, their AIC values, BIC values and maximum log-likelihood ratio values (LogLR) are shown in Table 5. The model with the smallest AIC value is the five-variable model containing the 5 variables with the smallest $p$-values, fhd+age+typ+tob+ldl. The model with the smallest BIC value is also this five-variable model. Since there are $p=9$ variables in the full model, the degrees of freedom of the $\chi^2$ distribution for calibrating the log-likelihood ratio is $10$. The $\chi^2_{1-\alpha, 10}$ quantiles defining the confidence regions (\ref{c_region}) associated with CMC$_{0.9}$, CMC$_{0.5}$ and CMC$_{0.1}$ are, respectively, 4.865,  9.341 and 15.987.  From the LogLR column in Table 5, we see that models with log-likelihood ratios below 4.865 (or in ${\cal C}_{0.9}$) are the last 5 models with 5 to 9 variables, so CMC$_{0.9}$ chooses the smallest model in this set, which is the model with 5 variables chosen by the AIC and BIC. Similarly, CMC$_{0.5}$ also chooses the same 5-variable model. On the other hand, models with log-likelihood ratios below 15.987 are the last 6 models with 4 to 9 variables, so  CMC$_{0.1}$ chooses the smallest model in this set of 6 models which is the 4-variable model consisting of the 4 variables with the smallest $p$-values, fhd+age+typ+tob. Although this model is different from the common choice of the other four criteria, it is worth considering as for this dataset the sample size $n=462$, which is 50 times larger than the number of variables $p=9$, and CMC$_{0.1}$ has been very accurate in our simulation study when the sample size is this large. % (see cases where $n\times m\geq 50p$ in Table 2).

McLeod and Xu (2020) analysed this dataset and obtained the above 5-variable model and 4-variable model, respectively, under two different BIC$_q$ criteria discussed in that paper. In Chapter 4 of their book, Hastie, Tibshirani and Friedman (2009) also analysed this dataset. They obtained a different 4-variable model containing fhd+age+tob+ldl using a backward selection method. Different model selection criteria may lead to different selections. The CMC criteria at different $\alpha$ levels are no exceptions, but the CMC provides a simple and unified framework to view the different selections through their log-likelihood ratios and associated $\alpha$ levels.

\section{Concluding remarks}

The CMC based on the log-likelihood ratio provides a family of criteria indexed by the $\alpha$ level for selecting regression models. It makes effective use of the null distribution of the likelihood ratio (\ref{a2}) for model selection. For general applications, we recommend CMC$_{0.5}$ as it has showed excellent accuracy in our simulation study, outperforming other criteria including AIC and BIC in most cases. With a parameter $\alpha$, it is easy for the CMC to adapt to special situations. There have been various efforts in finding finite sample adjustments for the AIC and BIC in order to improve their performance; see, for example, Hurvich and Tsai (1989), Broersen (2000) and Sclove (1987). The CMC does not need such adjustments. When the sample size is small or when there are strongly correlated predictor variables, we simply use CMC$_{\alpha}$ with a large $\alpha$ level, say $\alpha={0.9}$, to handle such special situations.

%The CMC as defined in (\ref{form1}) is for the best subset selection as the minimization is taken over the set ${\cal M}$ of all $2^p$ possible models, whereas the AIC and BIC may be applied to select a model from a subset ${\cal M}_s  \subset {\cal M}$. In situations where models of interest form such a subset ${\cal M}_s$, we simply replace the ${\cal M}$ in (\ref{form1}) with ${\cal M}_s$ so that the CMC can still select a model from ${\cal M}_s$. It is possible that ${\cal M}_s \cap {\cal C}_\alpha$ is empty, and in this case the CMC does not have a solution. This is a useful warning as it tells us that among the models in ${\cal M}_s$, none is acceptable at level $\alpha$ by the likelihood ratio test. As such, we should either refrain from selecting a model from ${\cal M}_s$, or consider lowering the $\alpha$ level to increase the set of plausible models so that ${\cal M}_s \cap {\cal C}_\alpha$ is not empty. In contrast, the AIC and BIC do not give such a warning and would select a model from ${\cal M}_s$ even if it contains no plausible models.

We have used the likelihood ratio test to define the set of plausible models. The score test and Wald test are asymptotically equivalent to the likelihood ratio test, and in principle they may also be used to define the set of plausible models for constructing the CMC. However, they are computationally more complicated than the likelihood ratio test. Further, one of the key argument used in establishing the lower bound (\ref{convg2}) for the CMC based on the likelihood ratio test is that  event $\{{\bm{\beta}^t}\in {\cal C}_{\alpha}\}$ implies $\{\hat{\bm{\beta}}_{j}^t \in {\cal C}_{\alpha}\}$. This argument would be invalid if other tests are used which will make the theoretical investigation of the CMC selection more difficult. Nevertheless, we plan to study Wald test based CMC to determine if it has theoretical advantages over the likelihood ratio test based CMC. In particular, letting $ {\cal C}_{\alpha}^W$ be the Wald test induced confidence region for $\bs{\beta}^t$, we hope to find a sequence of $\alpha_n\rightarrow 0$ such that $P(\bs{\beta}^t \in {\cal C}_{\alpha_n}^W) \rightarrow 1$ and 
\beq
\max_{\bs{\beta}\in {\cal C}_{\alpha_n}^W} \| \bs{\beta} -\hat{\bs{\beta}}\|_2 =o_p(1)  \nonumber
\eeq
uniformly for all ${\alpha_n}$ as $n\rightarrow \infty$. If such a sequence of $\alpha_n$ can be found, then we can show that the CMC defined by $\alpha_n$ is consistent, so it may have superior large sample accuracy than CMC$_{0.1}$. Wald test induced confidence region has an analytic expression which should be helpful in looking for such a sequence. The score test and likelihood ratio test induced confidence regions do not have this advantage.

%%%%%%%%%%%%%%%%%%%%%%%%%%%%%%%%%%%%%%%%%%%%%%%%%%%%%%%%%%%%%%%%%%%%%%%%%%%%%%%%%%%%%%%%%%%%%%%%%%%%%%%%%%%%%%%%%%%%%%%%%%%%
\vskip 14pt
%\noindent {\large\bf Supplementary Materials}

%The Supplementary Material is a self-contained document containing additional computational details and numerical examples, and %applications and discussions concerning the variability weighted average effect.
%\par
%%%%%%%%%%%%%%%%%%%%%%%%%%%%%%%%%%%%%%%%%%%%%%%%%%%%%%%%%%%%%%%%%%%%%%%%%%%%%%%%%%%%%%%%%%%%%%%%%%%%%%%%%%%%%%%%%%%%%%%%%%%%
%\vskip 14pt
%\noindent {\large\bf Acknowledgements}

%This work is supported by a research grant from the Southern University of Science and Technology of China and a Discovery Grant from the National Science and Engineering Research Council of Canada. We would like to thank two anonymous reviewers for comments that have led to improvements in this paper.

%\begin{ack}{ACKNOWLEDGEMENTS}

%\end{ack}


\begin{thebibliography}{}
% For English spelling, we follow the style of this dictionary

\renewcommand{\baselinestretch}{1.0}
{\small



\bibitem{r1}
{Akaike, H.} (1974).
A new look at the statistical model identification
\textit{IEEE Transactions on Automatic Control}, 19, 716--723.

\bibitem{r2}
Broersen, P. M. T. (2000). Finite sample criteria for autoregressive order
selection. \textit{IEEE Transactions on Signal Processing}, 48, 3550–3558.

\bibitem{r3}
Ding, J., Tarokh, V. and Yang, Y. (2018a).
Model Selection Techniques: An Overview.
\textit{ IEEE Signal Processing Magazine}, 35, 16--34.

\bibitem{r3*}
Ding, J., Tarokh, V. and Yang, Y. (2018b).
Bridging AIC and BIC: a new
criterion for autoregression. \textit{IEEE Transactions on Information Theory}, 64, 4024–4043.

\bibitem{r2**}
Fahrmeir, L., Kaufmann, H. (1985).
Consistency and asymptotic normality of the maximum likelihood estimator in generalized linear models.
\textit{Annals of Statistics}, 13, 342-368.


\bibitem{r3**}
Furnival, G. M. and Wilson, R. W., Jr. (1974).
Regressions by leaps and bounds.
\textit{Technometrics}, 16, 499–511.

\bibitem{r3***}
Gourieroux, C., Monfort, A. (1981).
Asymptotic properties of the maximum likelihood estimator in dichotomous logit models.
\textit{Journal of Econometrics}, 17, 83--97.

\bibitem{r9*}
Haberman, S. J.  (1977).  Maximum likelihood estimates in exponential response models. 
\textit{Annals of Statistics}, 5, 815--841.

\bibitem{r4}
{Hastie, T., Tibshirani, R.}, {Friedman, J.} (2009).
\textit{Elements of Statistical Learning: Data Mining, Inference and Predictions}. 2nd edition.
Springer Verlag, New York.
%\MR{838085}

\bibitem{r4*}
Hannan, E. J.  and Quinn, B. G.  (1979).  The determination of the order of an autoregression. 
\textit{Journal of Royal Statistical Society}, Series B, 41, 190–195.

\bibitem{r5} 
Hurvich, C. M. and Tsai, C. L. (1989).  Regression and time series model selection in small samples. \textit{Biometrika}, 76, 297–307. 

\bibitem{r6}
Kadane, J.~B. and Lazar, N.~A. (2004).
Methods and criteria for model selection,
\textit{Journal of the American Statistical Association},  99, 279--290.

\bibitem{r7}
McLeod, A. I., Xu, C. and Lai, Y. (2020). Package `bestglm'. An R package available at
https://cran.r-project.org.


\bibitem{r8}
McLeod, A. I. and Xu, C. (2020). `bestglm: Best Subset GLM'. Vignette for R package `bestglm' available at
http://www2.uaem.mx/r-mirror.


\bibitem{r8*}
Morgan J.A. and Tatar, J. F. (1972). Calculation of the Residual Sum of Squares for all Possible
Regressions. \textit{Technometrics}, 14, 317–325.


\bibitem{r20*}
Rao, C. R. and Wu, Y. H. (1989). A strongly consistent procedure for model selection in a regression problem. \textit{Biometrika},  76,  2, 369–374. 

\bibitem{r77}
Rousseauw, J., du Plessis, J., Benade, A., Jordaan, P., Kotze, J., Jooste, P.
and Ferreira, J. (1983). Coronary risk factor screening in three rural
communities, \textit{South African Medical Journal}, 64, 430–436.


\bibitem{r21}
{Schwarz, G. ~E.} (1978).
Estimating the dimension of a model,
\textit{Annals of Statistics}, 6, 461--464.

\bibitem{r22}
Sclove, S. L. (1987). Application of model-selection criteria to some problems in
multivariate analysis. \textit{Psychometrika}, 52, 333–343.

\bibitem{r33}
Tsao, M. (2021). A constrained minimum method for model selection. \textit{Stat}, e387. \\ Available at https://onlinelibrary.wiley.com/doi/10.1002/sta4.387




}

\end{thebibliography}
\end{document}